\documentclass[12pt,preprint]{aastex}

\def\fuse{{\it FUSE\/}}
\def\mfarcs{\hbox{$~\!\!^{\prime\prime}$}}
\def\hmol{H$_{\rm 2}$\/}

\shorttitle{D/H toward PN}
\shortauthors{Oliveira et al.}

\begin{document}

\title{Evidence for deuterium astration in the planetary nebula Sh\,2--216?\altaffilmark{1}}

\author{Cristina M. Oliveira\altaffilmark{2}, Pierre Chayer\altaffilmark{2,3}, H.~Warren~Moos\altaffilmark{2}, Jeffrey W. Kruk\altaffilmark{2}, and Thomas Rauch\altaffilmark{4}}
  
\altaffiltext{1}{Based on observations made with the NASA-CNES-CSA {\it Far Ultraviolet Spectroscopic Explorer}. \fuse~is operated for NASA by The Johns Hopkins University under NASA contract NAS5-32985.}
\altaffiltext{2}{Department of Physics and Astronomy, The Johns Hopkins University, 3400 N. Charles St., Baltimore MD 21218}
\altaffiltext{3}{Primary affiliation: Department of Physics and Astronomy, University of Victoria, P.O. Box 3055, Victoria, BC V8W 3P6, Canada.}
\altaffiltext{4}{Institut f\"{u}r Astronomie und Astrophysik, Universit\"{a}t T\"{u}bingen, Sand 1, 72076 T\"{u}bingen, Germany}

\begin{abstract} 

We present {\it FUSE} observations of the line of sight to WD\,0439$+$466 (LS\,V\,$+$46\,21), the central star of the old planetary nebula Sh\,2--216. The {\it FUSE} data shows absorption by many interstellar and stellar lines, in particular \ion{D}{1}, \hmol~($J$ = 0 -- 9), HD $J$ = 0 -- 1, and CO. Many other stellar and ISM lines are detected in the STIS E140M {\it HST} spectra of this sightline, which we use to determine $N$(\ion{H}{1}). We derive, for the neutral gas, D/H = (0.76 $\pm~^{0.12}_{0.11}$)$\times10^{-5}$, O/H = (0.89 $\pm~^{0.15}_{0.11}$)$\times10^{-4}$ and N/H = (3.24 $\pm~^{0.61}_{0.55}$)$\times10^{-5}$. We argue that most of the gas along this sightline is associated with the planetary nebula. The low D/H ratio is likely the result of this gas being processed through the star (astrated) but not mixed with the ISM. This would be the first time that the D/H ratio has been measured in predominantly astrated gas. The O/H and N/H ratios derived here are lower than typical values measured in other planetary nebulae likely due to unaccounted for ionization corrections.

\end{abstract}

\keywords{ISM: Abundances --- ISM: Evolution --- Ultraviolet: ISM --- Stars: Individual (WD\,0439$+$466, LS\,V\,$+$46\,21): ISM --- Planetary Nebulae (Sh\,2--216)}

\section{INTRODUCTION} 

Stars with intermediate mass (1--8 M$_{\odot}$) evolve through the asymptotic giant branch (AGB) experiencing large mass-loss rates, replenishing the interstellar medium with metals, on their way to become white dwarfs with $M$ $<$ 1.44M$_{\odot}$. When the stars become hot enough they photoionize the material ejected during the AGB phase, giving origin to a planetary nebula (PN). The gas in the PN reflects the stellar processing that occurred during the stellar life, for species such as helium, nitrogen, and carbon \citep{2006ApJ...651..898S}, while oxygen is not affected or only slightly destroyed during the AGB phase. Thus, the composition of PNe is an important tool for understanding the chemical evolution of the Galaxy. Deuterium is produced in appreciable quantities only in primordial big bang nucleosynthesis and easily destroyed in the interiors of stars \citep{1976Natur.263..198E,1971Ap&SS..14..179T}. Many studies have been performed through the years in order to determine the current gas-phase abundance of deuterium, which can be linked to the primordial abundance of deuterium through galactic chemical evolution models \citep[see for e.g.][and references therein]{2006ApJ...642..283O,2006MNRAS.369..295R}. These studies show that, despite infall, the abundance of D has decreased over time due to astration. The gas-phase D/H is constant throughout the Local Bubble (LB), (1.56 $\pm$ 0.04)$\times10^{-5}$ \citep{2004ApJ...609..838W}. Beyond the LB there is a large scatter in the measurements with values as low as (0.50 $\pm$ 0.16)$\times10^{-5}$ \citep[$\theta$ Car,][]{1992ApJS...83..261A} and as high as (2.18 $\pm~^{0.22}_{0.19}$)$\times10^{-5}$ \citep[$\gamma^2$ Vel,][]{2000ApJ...545..277S}. It has been proposed that the low gas-phase D/H values are due to deuterium being depleted in grains and that the total present day D/H in the local Galactic disk might be higher than 2.31$\times10^{-5}$ \citep{2004oee..symp..320D,2006ApJ...647.1106L}. In this work we present a study of D/H along the sightline to WD\,0439$+$466, the central star of the PN Sh\,2--216. We argue that the low gas-phase D/H ratio derived here, (0.76 $\pm~^{0.12}_{0.11}$)$\times10^{-5}$, is not due to deuterium being depleted in dust grains but is a consequence of the sightline being dominated by astrated, deuterium-poor gas associated with the PN. All uncertainties quoted are at the 1$\sigma$ level.

\section{WD\,0439$+$466 and its Planetary Nebula}

Sh\,2--216 is a large and old \citep[3--6$\times10^5$ yr,][]{1992MNRAS.259..315T,1999A&A...350..101N} low-surface brightness PN at a distance of 129 $\pm^{6}_{5}$ pc \citep[][trigonometric parallax]{2007AJ....133..631H} in the direction $l$ = 158.49$^\circ$, $b$ = $+$0.47$^\circ$. At this distance its apparent size of 1$\fdg$6 translates into a linear diameter of $\sim$3.6 pc. \citet{1981ApJ...245..131F} first proposed that Sh\,2--216 was a PN based on the strength of the [\ion{N}{2}] and [\ion{S}{2}] emission lines. A few years later \citet{1985ApJ...288..622R} identified WD\,0439$+$466 (DAO-type white dwarf) as a possible central candidate for the Sh\,2--216 PN. This identification was difficult because the central star is no longer located in the geometrical center of the nebula, probably due to the interaction of the PN with the ambient ISM. \citet{1985ApJ...288..622R} derived an expansion velocity $<$ 4 km s$^{-1}$ and a temperature $T$~= 9400 $\pm$ 1100 K for the nebula. From its distance and proper motion \citet{2004A&A...420..207K} determined that WD\,0439$+$466 left the geometrical center of the PN Sh\,2--216 about 45,000 years ago.




\section{{\it FUSE} and STIS OBSERVATIONS}

WD\,0439$+$466 has been observed by {\it FUSE} numerous times through all the apertures (programs M105, M107, and P104). We coadded all the calibrated data (CalFUSE version 3.0\footnote{The CalFUSE pipeline reference guide is available at http://fuse.pha.jhu.edu/analysis/pipeline\_reference.html}) for each aperture individually, following the procedures described in \citet{2006ApJ...642..283O}. The three resulting datasets have exposure times ranging from 10 ksec to 51.5 ksec. The {\it FUSE} data covers the spectral range 905 -- 1185 \AA~with a spectral resolution of $\sim$20 km s$^{-1}$ \citep{2000ApJ...538L...1M,2000ApJ...538L...7S}.


The {\it HST}/STIS data were obtained with the E140M echelle grating through the 0.2$\mfarcs\times0.2\mfarcs$ aperture and have a resolution of $\sim$5 km s$^{-1}$ (observations o64d03010 and o64d03020 with exposure times of 2.4 ksec and 3.0 ksec, respectively). The data, processed with CALSTIS version 2.18, were downloaded from the MAST archive at STScI and no further processing was applied.

\section{RESULTS AND DISCUSSION}

Column densities of several atomic and molecular species , determined from {\it FUSE} and {\it HST} data \citep[following the same procedures of][]{2006ApJ...642..283O}, are presented in Table \ref{columns}.

\begin{figure}
\begin{center}
\epsscale{0.70}
\rotatebox{90}{
\plotone{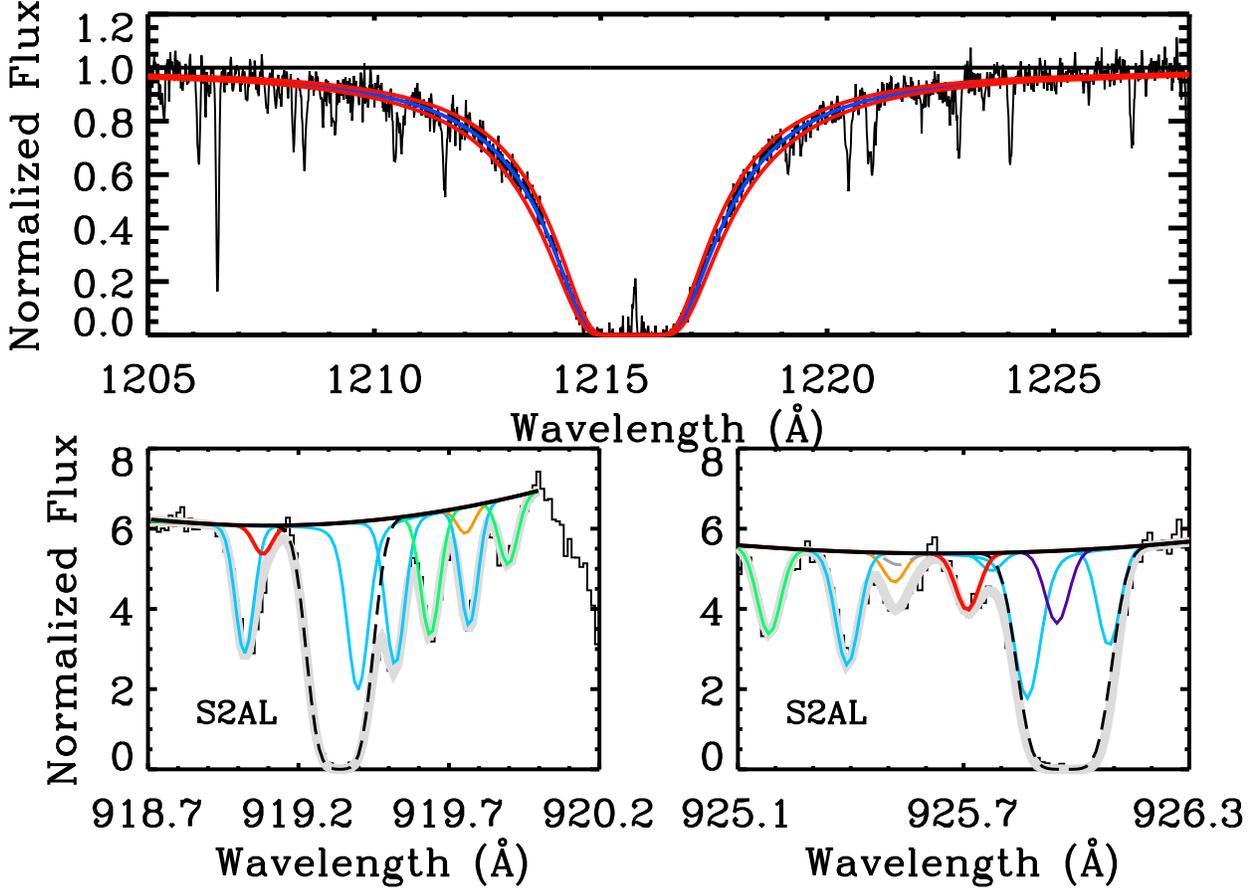}}
\caption{{\it Top Panel:} Fit to \ion{H}{1} Ly$\alpha$ (blue), obtained with a model with $T_{\rm eff}$ = 83,200 K, log $g$ = 6.74, and log (He/H) = $-$1.95 \citep{1999A&A...350..101N}, yielding log $N$(\ion{H}{1}) = 19.91. 1 $\sigma$ uncertainties of 0.06 dex are represented by red. {\it Bottom Panels:} Fits to some of the \ion{D}{1} lines used in the analysis (most and least blended \ion{D}{1} lines used, left and right panels, respectively). \ion{H}{1} is represented by a black dashed line, \ion{D}{1} in red, \ion{O}{1} in green, \ion{Fe}{2} in magenta, \hmol~in blue, HD in orange, and CO in dashed grey. The overall fit is represented by a thick grey line. The {\it FUSE} channel used in each plot is given at the bottom left. {\it S2AL} corresponds to LWRS SiC 2A.\label{hifit}}
\end{center}
\end{figure}

$N$(\ion{H}{1}) was determined from Ly$\alpha$ in conjunction with non-LTE stellar models \citep[with $T_{\rm eff}$ = 83200 $\pm$ 3300 K, log $g$ = 6.74 $\pm$ 0.19, and log (He/H) = $-$1.95 from][]{1999A&A...350..101N} in a manner similar to that of \citet{2006ApJ...642..283O}. Our fits yield log$N$(\ion{H}{1}) = 19.91 $\pm$ 0.06. Figure \ref{hifit} ({\it top panel}) presents the fit to the Ly$\alpha$ region (in blue). The 1$\sigma$ uncertainties of 0.06 dex (in red) combine the statistical uncertainties with the one associated with using different stellar models. Our $N$(\ion{H}{1}) is consistent with that obtained by \citet{rauch2006} using more recent models that also incorporate metals and that lead to $T_{\rm eff}$ = 95,000 K and log $g$ = 6.90. ($N$(H\,I) is not very sensitive to the $T_{\rm eff}$ of the stellar model because of the large breadth of the Ly$\alpha$ profile and the small stellar contribution to the profile.)

No stellar model was used to determine $N$(\ion{D}{1}). One-absorption-component fits to the $\lambda\lambda$919.1, 920.9 (blended with \hmol~$J$ = 6), 922.9 (blended with \hmol~$J$ = 3 and HD $J$ = 0), and 925.9 (minor blend with \hmol~$J$ = 7) \ion{D}{1} lines lead to $N$(\ion{D}{1}) = 15.09 $\pm$ 0.02. However the $\lambda$919.1 \ion{D}{1} line is not well fit, possibly due to a blend with an unknown stellar line. Removing this line from the fit yields $N$(\ion{D}{1}) = 15.00 $\pm$ 0.02. To take into account the different fits we adopt for this sightline $N$(\ion{D}{1}) = 15.05 $\pm$ 0.07. Figure \ref{hifit} ({\it bottom panels}) presents fits to some of the \ion{D}{1} lines used in the analysis. 

At the {\it FUSE} and STIS resolutions all the ISM lines display only a single absorption component with a common velocity, $v_{\rm helio}$ $\sim$ 6 km/s; hence in Table \ref{columns} we quote total sightline column densities.  However, the detection of several ionization stages for some of the species (e.g. \ion{C}{1}, \ion{C}{2}, and \ion{C}{4}) indicate that there must be at least three absorption components along this line of sight: a cool component traced by molecular hydrogen and \ion{C}{1} (amongst other species), a photoionized component traced by the high ionization species (\ion{C}{4}, \ion{N}{5}, and \ion{Si}{4}), and a warm component where the bulk of the species such as \ion{C}{2}, \ion{O}{1}, \ion{N}{1},\ion{S}{2}, \ion{Fe}{2}, etc. reside. The highly ionized species (no values of $N$ quoted in Table \ref{columns} because of blending with the stellar component) likely reside in the photoionized gas associated with the planetary nebula where optical emission lines of H$\alpha$ and \ion{O}{3}  have been detected \citep{1981ApJ...245..131F}. The LB contributes only log $N$(H) $\sim$ 19.3 \citep{1999A&A...346..785S} to the total neutral hydrogen column density, log $N$(H) = 20.30 (or $\sim$10\%), so most of the gas sampled here is beyond the LB. However, D/H derived below is only an upper limit to D/H in the PN because of D and H absorption by foreground gas.

We derive, for the neutral gas along this sightline, D/H = (0.76 $\pm~^{0.12}_{0.11}$)$\times10^{-5}$, N/H = (3.24 $\pm~^{0.61}_{0.53}$)$\times10^{-5}$, and O/H = (0.89 $\pm~^{0.13}_{0.11}$)$\times10^{-4}$ (see Table \ref{ratios}). Note that this sightline contains a significant amount of molecular gas. Therefore, the neutral gas ratios take $N$(\hmol) and $N$(HD) into account: D/H = ($N$(\ion{D}{1}) + $N$(HD))/($N$(\ion{H}{1}) + 2$N$(\hmol) + $N$(HD)), etc.

The D/H ratio is $\sim$2 times smaller than the LB value quoted above but comparable to D/H along other sightlines. However, the N/H and O/H ratios, $\sim$2.3 and $\sim$3.9 times smaller respectively than the typical ISM ratios of (7.5 $\pm$ 0.4)$\times10^{-5}$ \citep{1997ApJ...490L.103M} and (3.43 $\pm$ 0.15)$\times10^{-4}$ \citep{1998ApJ...493..222M}, are a strong indication that the gas sampled by this sightline is different from typical diffuse sightlines. It is unlikely that the O/H ratio is due to depletion into dust, as studies of the O/H ratio have found that oxygen is only mildly depleted in high mean density sightlines, with an average ratio of (2.84 $\pm$ 0.12)$\times10^{-4}$ \citep{2004ApJ...613.1037C}, significantly higher than the value derived here. Using equations 14--18 of \citet{1968ApJ...154...33P} in conjunction with the optical emission line data from Table 2 of \citet{1981ApJ...245..131F} we estimate, for the ionized gas, O/H $\sim$3.9$\times10^{-4}$ and N/H $\sim$7.3$\times10^{-5}$ for position C (closest to the central star) \citep[using $T_{\rm e}$ = 9400 K from][]{1985ApJ...288..622R}. These ratios, based on H$\alpha$, \ion{N}{2}, \ion{O}{2}, and \ion{O}{3} data, are more similar to O/H and N/H ratios typical of circular PNe \citep[$\sim$3.6$\times10^{-4}$ and $\sim$1.7$\times10^{-4}$, respectively, from][although a large scatter exists]{2006ApJ...651..898S}. Ionization corrections, not applied here, are then responsible for our low neutral-gas N/H and O/H ratios but can not explain our D/H ratio, as \ion{D}{1} and \ion{H}{1} have the same ionization fraction due to rapid charge exchange \citep{1973ApJ...184L.101B}. We note that the coupling of O and H ionization fractions is very strong unless there is appreciable ionization of O to higher stages \citep{2000ApJ...538L..81J} as is the case along this sightline. Other indications that most of the gas along this sightline is associated with the PN come from the molecular gas:
1) Molecules such as CO and H$_2$ exist in the envelopes of PNe. These molecules are distributed in a clumpy structure and in evolved PNe can survive only inside dense clumps or they would be rapidly photoionized and dispersed on time scales shorter than the age of the nebula \citep{1993IAUS..155..155T,2000ASPC..199..277H}. In addition, the fraction of H$_2$ along this sightline, $f_{\rm H_2}$ = 0.58 $\pm$ 0.06, is not typical of sightlines with such low $N$(H) and is one of the highest values in the literature for log $N$(H) $<$ 21.0 \citep[see e.g.][and references therein]{2006ApJ...636..891G}. Hence, the gas containing the CO and probably most of the \hmol~is likely in a dense clump in the envelope of the PN. 2) Transitions from high $J$ lines ($J$ = 7--9) and HD $J$ = 1 are not typical of diffuse cloud sightlines. These lines are formed in environments with radiation fields much more intense than the average interstellar field \citep{2006ARA&A..44..367S}, lending support to the idea that the molecular gas is close to the central star of the PN.

Although depletion of D into dust grains cannot be ruled out we think it is unlikely because: 1) Sh\,2--216 is much younger \citep[$\sim$ 0.3--0.6 Myr,][]{1992MNRAS.259..315T,1999A&A...350..101N} than the timescale for deuterium depletion into dust grains \citep[$\sim$2.3 Myr for D/H to decrease by 1/$e$,][]{2006ASPC..348...58D}. 2) Extreme enrichment of the grains relative to the gas, requires $T_{\rm dust}$ $<$ 90 K in order to explain the lowest D/H ratios observed \citep{2006ASPC..348...58D,2006ApJ...647.1106L} but dust in PNe is typically warmer \citep[90--150 K,][]{2000ApJ...532..384V}.

The low D/H, N/H, and O/H ratios combined with the detection of CO, high H$_2$ and HD $J$ lines, and the high $f_{\rm H_2}$ are consistent with the idea that the gas along this sightline is dominated by gas that is associated with the PN. In this scenario the low D/H ratio is not a consequence of depletion into dust grains (although this cannot be completely ruled out) but of the gas in the nebula being deuterium-poor as a result of astration. 

Even though astration has reduced the primordial abundance of deuterium to the lower values found today in the Galaxy, this would be the first time that we have seen directly the effects of astration in the D/H ratio of any single sightline \citep[see e.g.][and references therein, for a summary of D/H ratios and pssobile explanations for the low values]{2006ApJ...647.1106L}. This work raises the possibility that other low D/H ratios could be explained sometimes by the presence of astrated gas that has not yet mixed with the ambient gas, particularly for nearby sightlines where this effect is less likely to be averaged out.

\acknowledgments

We thank E. B. Jenkins and J. L. Linsky for useful discussions and suggestions. Based on observations made with the NASA/ESA Hubble Space Telescope, obtained from the Data Archive at the Space Telescope Science Institute, which is operated by the Association of Universities for Research in Astronomy, Inc. under NASA contract NAS5-26555. The profile fitting procedure, Owens.f, used in this work was developed by M. Lemoine and the French \fuse~Team.

\clearpage

\begin{deluxetable}{lclc}
\tablewidth{0pc}
\tablecaption{Adopted Column Densities (cm$^{-2}$)\label{columns}}
\tablehead{ 
\colhead{Species} & \colhead{Log $N$}  & \colhead{Species} & \colhead{Log $N$}}
\startdata
\ion{H}{1}	& 19.91 $\pm$ 0.06  	& \ion{Fe}{2}	& 14.11 $\pm$ 0.06	\\
\ion{D}{1}	& 15.05 $\pm$ 0.07	&\hmol~($J$ = 0) & 19.54 $\pm$ 0.05 \\
\ion{C}{1}	& 14.60 $\pm$ 0.10	&\hmol~($J$ = 1) & 19.37 $\pm$ 0.05 \\
C\,I$^{*}$	& 14.40 $\pm$ 0.06	&\hmol~($J$ = 2) & 18.6: \\
C\,I$^{**}$	& 14.02 $\pm$ 0.08	&\hmol~($J$ = 3) & 18.0:\\
\ion{C}{2}	& $\ge$14.40		&\hmol~($J$ = 4) & 16.7:\\
C\,II$^{*}$	& $\ge$14.25		&\hmol~($J$ = 5) & 16.5:\\
\ion{N}{1}	& 15.81 $\pm$ 0.07	&\hmol~($J$ = 6) & 15.00 $\pm$ 0.25 \\
\ion{N}{2}	& $\ge$14.17		&\hmol~($J$ = 7) & 14.85 $\pm$ 0.08 \\
\ion{O}{1}	& 16.25 $\pm$ 0.05	&\hmol~($J$ = 8) & 13.70 $\pm$ 0.10 \\
\ion{Al}{2}	& $\ge$12.58		&\hmol~($J$ = 9) & 13.30 $\pm$ 0.05 \\
\ion{Si}{2}	& $\ge$14.08		&\hmol	& 19.76 $\pm$ 0.04 \\
\ion{Si}{3}	& $\ge$12.67		& H & 20.30 $\pm$ 0.03 \\	
\ion{P}{2}	& $\ge$12.70		& HD~($J$ = 0)	& 14.54 $\pm$ 0.10\\
\ion{S}{1}	& 13.45 $\pm$ 0.08	& HD~($J$ = 1)	& 13.72 $\pm$ 0.05 \\
\ion{S}{2}	& $\ge$14.90		& HD & 14.60 $\pm$ 0.09\\
\ion{Cl}{1}	& $\ge$12.71		& D = \ion{D}{1} $+$ HD & 15.18 $\pm$ 0.06\\
\ion{Ar}{1}	& $\ge$13.53		& CO	& 13.78 $\pm$ 0.07\\
\enddata
\tablecomments{Quoted uncertainties are 1$\sigma$. For \hmol~$J$ = 2--5 the column densities are uncertain; these are flagged with ':'. We use $N$(\hmol) $\simeq$ $N$($J$ = 0) $+$ $N$ ($J$ = 1) since these levels contain most of the \hmol~along the sightline. $N$(H) = $N$(\ion{H}{1}) + 2$N$(\hmol) + $N$(HD) }
\end{deluxetable}

\clearpage

\begin{deluxetable}{lclc}
\tablewidth{0pc}
\tablecaption{Ratios of Column Densities \label{ratios}}
\tablehead{ 
\colhead{Species} & \colhead{Ratio} &  \colhead{Species} & \colhead{Ratio}}
\startdata
D/H	& (0.76 $\pm~^{0.12}_{0.11}$)$\times10^{-5}$  & N/O & 0.36 $\pm~^{0.08}_{0.07}$	\\
\ion{D}{1}/\ion{H}{1}& (1.38 $\pm~^{0.32}_{0.27}$)$\times10^{-5}$ & HD/2\hmol & (3.39 $\pm~^{0.82}_{0.67}$)$\times10^{-6}$\\
O/H 	& (0.89 $\pm~^{0.13}_{0.11}$)$\times10^{-4}$  & HD/D & 0.26 $\pm~^{0.07}_{0.06}$ \\
N/H 	& (3.24 $\pm~^{0.61}_{0.53}$)$\times10^{-5}$  & CO/\hmol  & (1.05 $\pm~^{0.21}_{0.18}$)$\times10^{-6}$ \\
Fe/H 	& (0.65 $\pm~^{0.11}_{0.09}$)$\times10^{-6}$  & $f_{\rm H_2}$ & 0.58 $\pm$ 0.06 \\
D/O 	& (8.51 $\pm~^{1.63}_{1.44}$)$\times10^{-2}$  & $\langle n_{\rm H}\rangle$ (cm$^{-3}$)	 & 0.54 \\
\enddata
\tablecomments{Quoted uncertainties are 1$\sigma$. $N$(D) = $N$(\ion{D}{1}) + $N$(HD), $N$(H) = $N$(\ion{H}{1}) + 2$\times N$(\hmol) + $N$(HD), and $f_{\rm H_2}$ = 2$\times N$(\hmol)/$N$(H). For the ratios above we assume $N$(O) $\approx$ $N$(\ion{O}{1}), $N$(N) $\approx$ $N$(\ion{N}{1}), and $N$(Fe) $\approx$ $N$(\ion{Fe}{2}).}
\end{deluxetable}

\clearpage

\end{document}